# Detecting Aortic Valve Opening and Closing from Distal Body Vibrations

Andrew D. Wiens, *Student Member, IEEE*, Ann Johnson, and Omer T. Inan, *Senior Member, IEEE*

*Abstract— Objective:* Proximal and whole-body vibrations are well studied in seismocardiography and ballistocardiography, yet distal vibrations are still poorly understood. In this paper we develop two methods to measure aortic valve opening (AVO) and closing (AVC) from distal vibrations. *Methods:* AVO and AVC were detected for each heartbeat with accelerometers on the upper arm (A), wrist (W), and knee (K) of 22 consenting adults following isometric exercise. Exercise-induced changes were recorded with impedance cardiography, and nine-beat ensemble averaging was applied. Our first method, FilterBCG, detects peaks in distal vibrations after filtering with individually-tuned bandpass filters while RidgeBCG uses ridge regression to estimate AVO and AVC without peaks. Pseudocode is provided. *Results:* In agreement with recent studies, we did not find peaks at AVO and AVC in distal vibrations, and the conventional R-J interval method from the literature also correlated poorly with AVO ($r^2 = 0.22$ A, 0.14 W, 0.12 K). Interestingly, distal vibrations filtered with FilterBCG resembled seismocardiogram signals and yielded reliable peaks at AVO ($r^2 = 0.95$ A, 0.94 W, 0.77 K) and AVC ($r^2 = 0.92$ A, 0.89 W, 0.68 K). *Conclusion:* FilterBCG measures AVO and AVC accurately from arm, wrist, and knee vibrations, and it outperforms R-J intervals and RidgeBCG. *Significance:* To our knowledge, this study is the first to measure AVC accurately from distal vibrations. Finally, AVO timing is needed to assess cardiovascular disease risk with pulse wave velocity (PWV) and for cuff-less diastolic blood pressure measurement via aortic pulse-transit time (PTT).

## I. Introduction

THE timings of aortic valve opening (AVO) and closing (AVC) events reveal important physiologic information about the performance of the heart. When the occurrence of the electrocardiogram (ECG) R-wave is also known, the three systolic time intervals – pre-ejection period (PEP), left ventricular ejection time (LVET), and total electromechanical systole (QS2) – can be found directly from AVO and AVC. These time intervals are metrics of cardiac function that are clinically useful and can be measured noninvasively [1]–[5]. For example, systolic time intervals (STI) have been used to

This work was supported by the National Institutes of Health, National Institute of Biomedical Imaging and Bioengineering, grant number 1U01EB018818-01.

A. D. Wiens, A. Johnson, and O. T. Inan are with the Georgia Institute of Technology, Atlanta, GA, USA. (correspondence e-mail: andrew.wiens@gatech.edu).

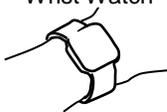

Fig. 1. Comparison of wearable pulse transit time (PTT) measurement modalities that can potentially be enabled by the technologies described in this paper. Distal aortic valve opening detection using ballistocardiogram (BCG) signals can even allow, for the first time, the measurement of *aortic* PTT in a knee brace form factor.

successfully quantify left ventricular (LV) function of healthy and diseased myocardia, and reduced LV function was found to increase PEP and decrease LVET [2], [3], [5]. More recently, STI from tissue Doppler imaging predicted rehospitalization of chronic heart failure patients [1]. Other noninvasive methods for measuring STI have been studied. Classically, STI are determined from an ECG, carotid pulse tracing, and phonocardiogram [3]. More recently, photoplethysmography and echocardiography have also been used to measure STI [6]. Another modern method, impedance cardiography (ICG), injects a small current through the upper body to measure changes in thoracic blood volume [7]–[9]. ICG has been used to distinguish healthy and impaired LV function in cases of chronic heart failure [4].

However, a major drawback of these methods is that each requires a proximal measurement to determine PEP. Applied distally, they cannot measure PEP because AVO and AVC at the periphery necessarily include a propagation delay [6]. This is an important problem to address because solving it could enable devices to measure blood pressure accurately from aortic pulse transit time (PTT) or to assess cardiovascular disease (CVD) risk from aortic pulse wave velocity (PWV). PTT, the time interval during which a pressure wave travels between two locations in the arterial tree, allows detection of beat-by-beat changes in blood pressure without a pressure cuff [10], [11]. Beat-by-beat sensitivity and lack of a pressure cuff are two significant advantages over today's standard blood pressure methods, such as auscultation and oscillometry [12], [13]. Although systolic blood pressure may be acquired with pulse arrival time without knowledge of any STI, recent studies indicate that PEP must be known in order to accurately resolve diastolic blood pressure [14], [15]. Also, aortic PWV, which is inversely proportional to aortic PTT and can be determined similarly once PEP is known, has been discovered as a CVD risk factor of high predictive power [16]–[20] and may be superior to any other predictor [21]. Thus, a novel methodology must be developed before any distally-worn PTT or PWV device can be implemented.

We believe that ballistocardiography (BCG) could meet this need. BCG is a time series of small vibrations of the body caused by the beating heart [22]. Recently, we conducted the first intensive study of BCG applied to PTT and found that BCG could be used as a reference for PEP [23]. Conveniently, BCG can be measured with a wide variety of sensors [24]–[30]. When measuring whole-body vibrations with a modified weighing scale, the time interval from the ECG R-wave to the BCG J-wave, or R-J interval, was found to correlate ($r^2$ = 0.86) with PEP [31]. This method has become the standard method for measuring PEP with any kind of BCG sensor [32]. In a different body of research, the vibration sensor is placed proximally on the sternum, and the resulting signal is called a seismocardiogram (SCG). The resulting time series contain peaks at AVO and AVC [33]–[41].

However, robust methods for determining AVO and AVC when the vibration sensor is worn distally are lacking [26], [28], [32]. In previous work, we found that naïve use of R-J intervals in distal BCG time series produced poor results, although R-J intervals have been widely used this way [32]. To address this problem, we developed a system identification method to adapt the waveform from a distal accelerometer to that of a weighing scale with a filter [42]. A main goal of that work was to improve PEP measurements from the standard R-J interval method of [31]. Although we found that our system identification method reduced the standard deviation of PEP error by an average of 36.4% in fifteen subjects, the method is limited by its need for a simultaneous reference BCG to tune the filter. Recently, we showed that a bandpass filter trained from ground truth AVO measurements improved PEP in recordings without the need for a ground-truth BCG reference [43]. However, that method uses the R-J interval technique, which can only measure PEP.

The need for distal capability is not always immediately obvious. While capturing an SCG with a device worn on the sternum is physically feasible in almost any situation where measurement from the arm, wrist, or knee would also be possible, there are two major situations where placement on the sternum would not be ideal. First, unlike medical devices or lab equipment, consumer devices must be both familiar and convenient to the user in order to gain widespread adoption. There is no existing ubiquitous device we can think of that is placed on the sternum. However, wrist watches, exercise armbands, and knee braces are all very common consumer accessories, and distal methods could be directly applied to such devices. Second, measurement of pulse-transit time requires knowing when the arterial pulse wave arrives distally at the end of a major blood vessel, such as the aorta, as well as the proximal timing of the heart. A single device located distally at the end of such a blood vessel that measures both the pulse wave arrival and the aortic valve opening (AVO) would be much more convenient – and less expensive – than two separate devices. In the latter case, one device would have to be placed at the sternum to capture an SCG along with another device worn at the end of the blood vessel, and these devices would have to be connected either wirelessly or with a cable. This would introduce unnecessary cost and complexity over a single-device solution. Fig. 1 summarizes three possible measurement systems that could leverage the approaches presented in this paper to enable wearable PTT, and even wearable aortic PTT, measurement.

The goal of this work is to develop two new robust methods for measuring aortic valve timing from distal vibrations and to demonstrate the impact of sensor placement on measurement quality. We explain how to measure AVO and AVC – and thus LVET and PTT – in time series from an accelerometer worn distally on the body. To our knowledge, FilterBCG is the first method to accurately detect AVC in distal vibrations. Training either method on a subject requires one recording with reference AVO and AVC measurements. Also, the methods are agnostic to accelerometer orientation until it is trained; the user can simply place the sensor on any convenient place on the upper body. Finally, we demonstrate these techniques on recordings from accelerometers on the upper arm, wrist, and knee during post-exercise recovery. This work should be valuable to anyone seeking proximal time intervals at distal locations on the body, such as researchers of BCG, PTT, and PWV.

## II. Materials and Methods

This study was approved by the Institutional Review Board of Georgia Institute of Technology (Protocol H13512). Twenty-two healthy adults (8 females, 14 males), whose detailed demographics appear in Table 1, participated in this study. The age distribution of the participants was 23.9 ± 7.5 years, and the weight distribution was 70.4 ± 13.8 kg. For each subject, blood pressure was measured (BP786, Omron Corporation, Kyoto, Japan), then the subject performed an isometric exercise (wall sit) for two minutes or until exhaustion, whichever occurred first. Each subject then stood vertically, motionless but relaxed, for a five-minute recovery period immediately after the exercise during which the time series were recorded. Five minutes was chosen as the recording length because it was observed that subjects nearly reached their resting PEP by that time. The purpose of the protocol was to perturb the system via the exercise intervention to acutely reduce PEP and LVET (due to increased sympathetic tone during exercise) compared to the resting values, and observe the return of both parameters back to resting values during recovery.

## A. Equipment

A commercial data acquisition system (MP150, BIOPAC Systems, Goleta, CA) was used to capture all physiologic signals with a standard PC. The physiologic instruments consisted of an ECG configured for a modified Lead II measurement (BN-EL50 wireless ECG, BIOPAC Systems), an 8-lead ICG providing a first derivative (dZ/dt) time series containing the AVO and AVC ground truth (NICO100C, BIOPAC Systems, Goleta, CA), and three triaxial accelerometers (3 x 356A32, PCB Piezotronics, Depew, NY) that were powered and amplified 100x with three Integrated Electronic Piezoelectric (IEPE) signal conditioners (482C05, PCB Piezotronics, Depew, NY). Impedance cardiography was chosen because it is easier to acquire than echocardiograms, which require a trained technician; ICG has been shown to measure AVO and AVC well in healthy subjects when compared to echocardiography [8], [9]; and ICG has been used previously in similar studies [27], [29], [31], [43]–[45]. All signals were sampled simultaneously at $f_s$ = 2 kHz and saved for offline processing in MATLAB (MathWorks, Natick, MA). The accelerometers were chosen for their low noise properties as described in [46] and were attached laterally to the upper arm, wrist, and knee with kinesiology tape wrapped around each limb (Kinesio Tex, Kinesio, Albuquerque, NM). Since each accelerometer consisted of three axes, and simultaneous ECG and ICG recordings were also obtained, a total of 11 synchronized time series were recorded.

## B. Statistical Methods

Standard methods from BCG and ICG literature were applied. The locations of the ECG R-waves were determined with a peak detector and used to segment each BCG and ICG recording into individual beats. Then, every nine beats (N = 9) in each time series were averaged sample-by-sample to generate ensemble averages as in [24], and the B-wave (AVO) in each dZ/dt beat was determined using a method we described in [43]. The dZ/dt X-wave (AVC) was determined as the lowest trough – the standard method in impedance cardiography [47] – from 150–330 ms after the B-wave. This range corresponds approximately to the expected range of LVET for healthy subjects [48], and it covered this dataset. Each detected B- and X-wave (AVO and AVC) was then inspected manually beat-by-beat to ensure that the ground-truth was accurate.

First, a modified version of the standard method to obtain AVO from R-J intervals via regression [31] that we first used in a recent pilot study [43] was applied to the BCG. The difference between the R-J interval method in [31] and that of this work is that individual regressions were found from the R-J intervals of each subject to ground-truth PEP after bandpass filtering. Each beat was filtered with a 6th-order Butterworth infinite impulse response (IIR) filter in the forward and reverse direction to achieve zero phase delay. Then, the filter cutoff frequencies, axes (x, y, z), and polarities (non-inverted or inverted) of the sensors that resulted in the lowest root mean square error (RMSE) after linear regression with ground-truth PEP were chosen for each subject and location (upper arm, wrist, and knee). The result was a single-axis filtered BCG for each location that contained the best possible J-wave measurement.

After determining R-J intervals, a new method called FilterBCG was applied to determine AVO and AVC using a peak-detection approach similar to standard methods in SCG [33], [37], [40], [41]. First, subject-specific IIR bandpass filters (6th-order Butterworth IIR, forward and backward) were trained to produce peaks between 1–150 ms and 200–475 ms after the ECG R-wave that resulted in the lowest RMSE after regression with ground-truth AVO and AVC, respectively. In other words, for each wearable location (upper arm, wrist, knee) and event (AVO or AVC) of each subject, the bandpass cutoff frequencies, feature type (peak, trough), and accelerometer axis (x, y, z) that produced the lowest RMSE after regression were selected.

Figure 2 shows an example of the filtered R-J interval and FilterBCG methods applied to vibrations of the wrist for one heartbeat aligned with several other time series. Trained IIR bandpass filters result in J-waves, AVO peaks, and AVC peaks that are more distinct than those appearing in the raw BCG waveform. These peaks are then regressed linearly to ground-truth AVO (or PEP), AVO, and AVC, respectively. Pseudocode for FilterBCG is shown in Figure 3 where the inputs are the result from this optimization step.

TABLE I
SUBJECT DEMOGRAPHICS

| # | Sex | Age | Weight | Height | Systolic BP* | Diastolic BP* |
|---|-----|-----|--------|--------|--------------|---------------|
| 0 | F | 19 yrs | 59 kg | 165 cm | 97 mmHg | 75 mmHg |
| 1 | F | 19 | 68 | 170 | 93 | 59 |
| 2 | M | 20 | 57 | 170 | 100 | 64 |
| 3 | F | 18 | 66 | 175 | 100 | 67 |
| 4 | M | 33 | 79 | 185 | 144 | 97 |
| 5 | F | 23 | 52 | 160 | 96 | 68 |
| 6 | M | 25 | 102 | 183 | 146 | 89 |
| 7 | M | 33 | 78 | 180 | 122 | 70 |
| 8 | F | 19 | 59 | 173 | 103 | 60 |
| 9 | M | 24 | 68 | 178 | 121 | 69 |
| 10 | M | 23 | 88 | 191 | 126 | 76 |
| 11 | M | 20 | 61 | 180 | 117 | 66 |
| 12 | M | 31 | 86 | 175 | 131 | 75 |
| 13 | F | 49 | 70 | 163 | 118 | 74 |
| 14 | F | 16 | 53 | 170 | 120 | 76 |
| 15 | M | 31 | 83 | 178 | 100 | 65 |
| 16 | F | 23 | 61 | 165 | 104 | 70 |
| 17 | M | 19 | 57 | 165 | 93 | 64 |
| 18 | M | 19 | 57 | 175 | 113 | 82 |
| 19 | M | 22 | 76 | 173 | 104 | 65 |
| 20 | M | 20 | 88 | 188 | 114 | 63 |
| 21 | M | 20 | 82 | 193 | 113 | 71 |

*BP, blood pressure.

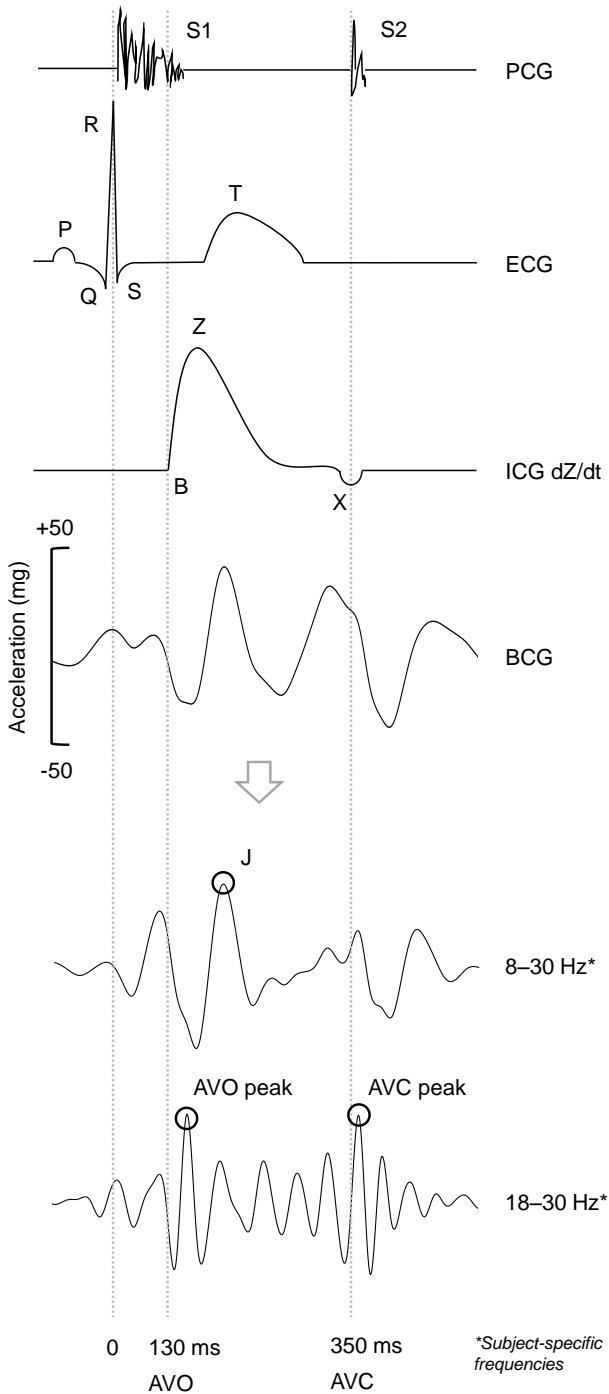

Fig. 2. Bandpass filtering enhances peaks in a distal vibration heartbeat from the wrist. Time series of a phonocardiogram (PCG), electrocardiogram (ECG), and impedance cardiogram 1st derivative (ICG dZ/dt) are shown as temporal references to the wrist ballistocardiogram (BCG). Three bandpass filters are trained on each BCG sensor: one for filtered R-J interval detection and two for AVO and AVC detection with FilterBCG. In this example, an 8–30 Hz passband produced the best R-J intervals, and 18–30 Hz provided the best AVO and AVC peaks for FilterBCG. When applied to a distal BCG recording, FilterBCG results in a BCG signal resembling an SCG that has two peaks coinciding with AVO and AVC.

Finally, a second new method based on ridge regression called RidgeBCG was applied. Unlike the R-J interval and FilterBCG methods, RidgeBCG does not choose peaks from the vibrations; rather, RidgeBCG determines AVO and AVC from all time-domain samples in downsampled beats that are

```
Algorithm FilterBCG
 1: procedure GETVALVEEVENTS(f_lo, f_hi, useMins, type)
 2:   win ← 1—150 ms if type is AVO else 200—475 ms.
 3:   dZdtBeats ← ensembleAverage(dZdt, Rwaves, N).
 4:   bcg ← filter(vibrationTimeSeries, bandpass(f_lo, f_hi)).
 5:   bcgBeats ← ensembleAverage(bcg, Rwaves, N).
 6:   groundTruth ← findWaves(dZdtBeats, win).
 7:   if useMins
 8:     AVevents ← findMinTroughInEachBeat(bcgBeats, win).
 9:   else
10:     AVevents ← findMaxPeakInEachBeat(bcgBeats, win).
11:   end
```

Fig. 3. A pseudocode representation of FilterBCG. The inputs specify the passband frequencies, whether to search for minima or maxima, and the type of aortic valve event (opening or closing).

located 1–150 ms and 200–475 ms after the ECG R-wave, respectively. RidgeBCG is implemented as follows: Before ensemble averaging, the BCG time series are filtered (bandpass, passband: 0.8–25 Hz, stopband: 35 Hz, 120 dB attenuation). The passband must include the cardiogenic frequencies of the BCG. Next, the vibrations are sliced at the ECG R-waves and ensemble averaged. The samples in 1–150 ms and 200–475 ms after the ECG R-wave in each beat are then taken as the AVO and AVC features, respectively; these are the same windows used in ICG and FilterBCG processing. The AVO and AVC features of each beat are normalized by subtracting the mean and dividing by the standard deviation of each beat, and the features are then decimated (downsampled) to a lower sampling rate $f_s$. In this study, the original $f_s$ was 2 kHz, the decimation was 28 samples, and the new $f_s$ was 71.4 Hz. This resulted in 11 and 20 samples for AVO and AVC, respectively.

This $f_s$ was appropriately chosen because the highest frequencies in the BCG signal occur below ~35 Hz. In practice, decimation should be applied such that the new $f_s \geq$ 70 Hz. As an alternative, the accelerometer can be sampled at or slightly above 70 Hz if band-limited to $\leq 0.5\, f_s$ prior to digitization. In such a case, decimation can be omitted. This approach would be particularly suitable for low-power devices.

The amplitudes of the decimated samples from each beat are then organized into vectors called feature vectors. This process is shown in Figure 4. These vectors are then organized into AVO and AVC feature matrices where each row is a normalized AVO or AVC feature vector from one beat. Finally, ridge regression [49] is applied using these matrices and AVO and AVC ground-truth measurements. Pseudocode for RidgeBCG appears in Figure 5.

The objective of ridge regression is to find the optimal weights, in the L2-norm regularized least-squares sense, which can obtain AVO or AVC via a linear combination of the entries in each feature vector. Ridge regression has one free parameter, $\lambda$, which is a scalar that determines the trade-off between bias and variance of the resulting measurement. Since the number of weights is equal to the number of samples in the feature vector, the weights will tend to overfit the data. The ridge parameter $\lambda$ prevents such overfitting. In this work, we used the average error of twenty repetitions of two-fold cross

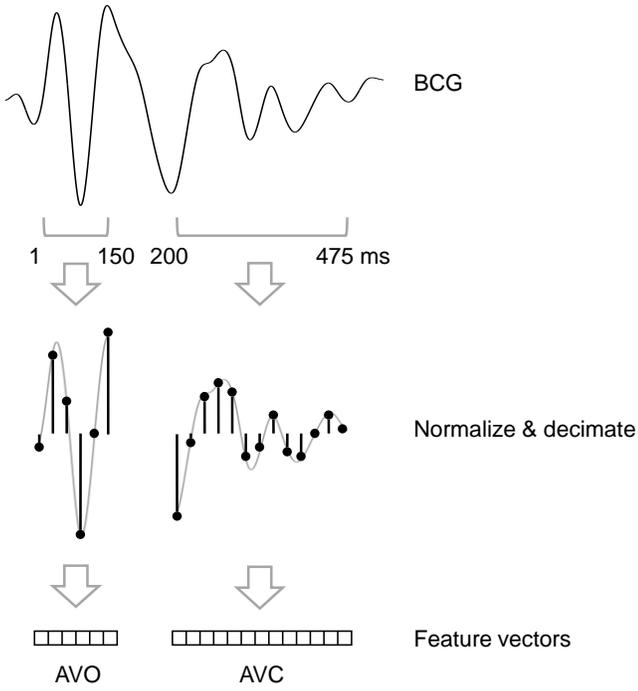

Fig. 4. RidgeBCG feature extraction. Rather than determining AVO and AVC from specific features, such as peaks, RidgeBCG measures AVO and AVC by downsampling the BCG heartbeats and applying ridge regression to feature vectors. These vectors contain the amplitudes of all samples in two predefined windows which cover the expected range of AVO and AVC.

---

**Algorithm** RidgeBCG

1: *eventType* ← {1–150ms,'B'} **if** AVO **else** {200–475ms,'X'}.
2: **procedure** TRAINRIDGEBCG
3:    *bcg* ← filter(*vibrationTimeSeries*, bandpass(0.8–25 Hz)).
4:    *bcgBeats* ← ensembleAverage(*bcg*, *Rwaves*, *N*).
5:    *dZdtBeats* ← ensembleAverage(*dZdt*, *Rwaves*, *N*).
6:    *featMatrix* ← new empty matrix.
7:    **for** *beat* **in** *bcgBeats*
8:       *featVec* ← getFeatureVector(*beat*, *eventType*).
9:       *featMatrix*.appendRow(*featVec*).
10:   **end**
11:   *groundTruth* ← findWaves(*dZdtBeats*, *eventType*).
12:   *b* ← ridgeRegression(*groundTruth*, *featMatrix*, λ).
13: **procedure** USERIDGEBCG
14:   *bcg* ← filter(*recentVibrations*, bandpass(0.8–25 Hz)).
15:   *Rw* ← the most recent *N*+1 Rwaves.
16:   *bcgBeat* ← ensembleAverage(*bcg*, *Rw*, *N*).
17:   *featVec* ← getFeatureVector(*bcgBeat*, *eventType*).
18:   *event* ← dotProduct(*featVec*, *b*).

Fig. 5. Pseudocode for RidgeBCG. RidgeBCG is trained first by applying ridge regression to a training set to obtain a vector of weights. After this step, AVO or AVC are measured as the dot, or inner, product of these weights with the feature vector of any heartbeat. An entry of 1 may need to be inserted at the beginning of each vector to accommodate a y-intercept.

---

validation (20 x 2-fold CV) [50]. The error metric used was the RMSE of AVO and AVC from RidgeBCG with respect to the ICG-derived ground truth AVO and AVC. The best λ of the best axis on each accelerometer was then chosen for each subject.

Since the exercise intervention was performed once per subject, one dataset was used for both training and testing. However, in real use, the training set should be obtained before using RidgeBCG. Thus, in order to determine results that could be expected in actual use, 20 x 2-fold CV was performed a second time to train the method on half of the beats and test the method on the others, where the ridge regression parameter λ was determined by a previous 20 x 2-fold CV training step as described earlier. In this paper, the results we present for RidgeBCG are the mean across all folds in this second CV step. Thus, our results are a conservative estimate of RidgeBCG's performance where overfitting was properly minimized. While CV was needed with RidgeBCG to prevent overfitting because RidgeBCG has 11 and 20 free parameters for AVO and AVC, CV was not used for FilterBCG or R-J intervals because the bandpass filter was trained on only two parameters – the upper and lower cutoff frequencies – which does not tend to overfit the training data.

### III. RESULTS AND DISCUSSION

The correlations between BCG- and ICG-derived AVO and AVC for all subjects and methods are shown in Figure 6 and Figure 7. Additionally, Figure 8 quantifies the effect of ensemble averaging on the FilterBCG correlations. The latter plot can be used to pick a suitable ensemble average length in future work. A total of 9648 heart beats from 22 subjects were processed, which resulted in 1072 beats after nine-beat ensemble averaging. For AVO and AVC from every location except the knee, FilterBCG resulted in better correlations than R-J intervals, filtered R-J intervals, or RidgeBCG. Filtered R-J intervals was the best method at the knee. This could be due to more attenuation of higher frequencies used in FilterBCG, since the R-J interval relies on lower frequencies to determine PEP.

Overall, the knee was worse than the upper arm and the wrist. As shown in Table 2, the within-subject RMSE differences of the wrist and upper arm AVO and AVC were not significantly significant in paired t-tests, but the wrist and knee were significantly different ($p < 0.05$). We believe that the knee was inferior to the wrist and upper arm due to higher attenuation of cardiogenic vibrations. Since the feet are grounded to a solid surface while standing, attenuation should increase, and thus the BCG signal-to-noise ratio (SNR) should decrease, as the sensor is worn lower on the body. This could be a question for further study.

Table 2 also shows that the difference between FilterBCG and filtered R-J intervals was not statistically significant. However, the trained bandpass filter significantly reduced the RMSE of each subject in both the FilterBCG and filtered R-J interval methods. The trained bandpass filter introduced in this paper is therefore a significant improvement over the unfiltered R-J intervals method, which is the current standard.

In the correlation summary in Figure 7, the best axis was used. In 41 out of 66 (62%) of the sensors in the study for 22 subjects and three wearable locations on the body, the best axis was the head-to-foot axis. To better understand the impact of the physical orientation of the sensor, FilterBCG was applied to each of the other two axes as well. In this case, the correlations were lower but still strong (AVO: $r^2$ = 0.82 A, 0.74 W, 0.49 K; AVC: $r^2$ = 0.68 A, 0.58 W, 0.45 K). The difference between the best axis and the other axes was statistically significant in all cases, and the p-values are shown in Table 2. However, the results were still acceptable from the suboptimal axes, especially for AVO from the upper arm ($r^2$ = 0.82). This suggests that it may be possible to obtain acceptable results even when the accelerometer is placed haphazardly on the body. This is important because single-axis accelerometers can be lower in cost and power consumption than multi-axis variants, and the ability to place a BCG device on the body without great care also simplifies the technique for the end-user. Nevertheless, the optimal axis seems to vary subject-to-subject and may also depend on precisely how the sensor is placed. It therefore appears that better results can be obtained with a multi-axis accelerometer in general. Developing better methods to combine several BCG axes could be a topic of future study. Also, increasing the number of axes beyond three, for example with a gyroscope to measure angular motion, could also be a topic for future study.

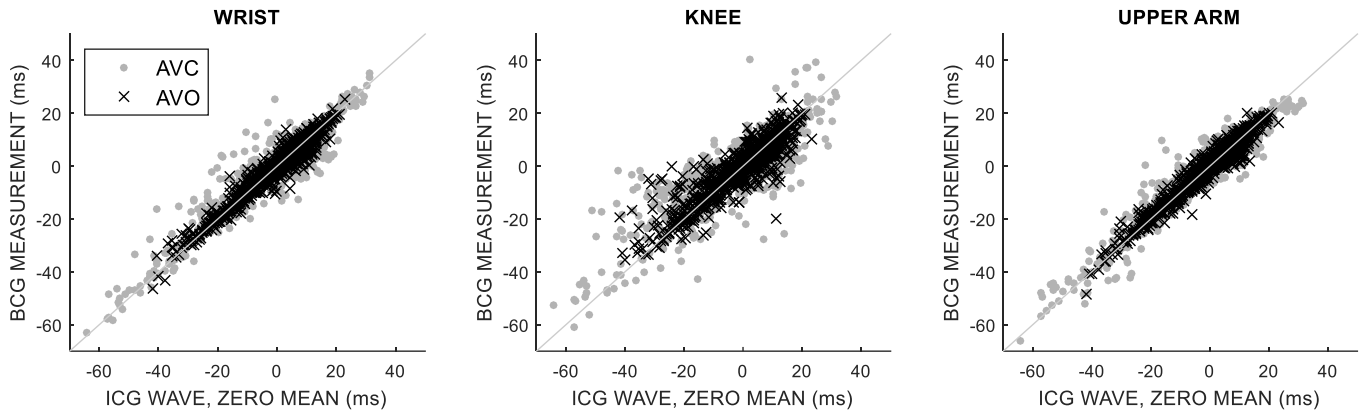

Fig. 6. Scatterplots of FilterBCG measurements versus an ICG ground-truth for the three accelerometers worn on the body after training. The ICG and vibration time series were nine-beat ensemble averaged, and each subject's ICG measurements were centered about zero before FilterBCG to remove differences between the subjects' AVO and AVC baselines. When intra-subject means are not subtracted, the resulting correlations tend to be optimistic because most of the variance is represented by inter-subject spread rather than the method of measurement.

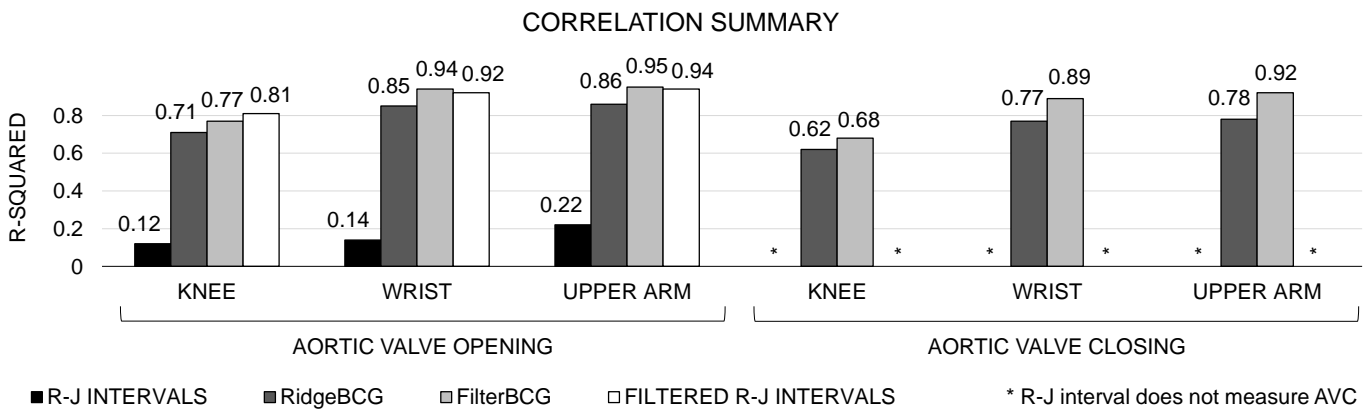

Fig. 7. Correlations of BCG to ICG for all methods after nine-beat ensemble averaging. R-J intervals do not measure AVC and thus do not appear on the right. A key contribution of this paper is the finding that distal wearable ballistocardiograms can measure AVC after bandpass filtering. In the case of the wrist and upper arm, such measurements are quite accurate. FilterBCG exceeded other methods in this study. Although filtered R-J intervals did have a higher correlation than FilterBCG at the knee, the differences between FilterBCG and filtered R-J intervals were not statistically significant as shown in Table 2.

The principal limitations of FilterBCG and RidgeBCG include subject-specific training and requiring the subject to reduce body movements during measurements. Subject-specific training on time series during which AVO and AVC are changing is a major inconvenience for users. The vast majority of users of wearable devices will have neither ground-truth measurements nor the time to perform in-depth calibrations for their devices. Future work should focus on reducing or eliminating the training step. The latter limitation of reducing body movement applies to any method relying on small body vibrations, including all techniques in the ballistocardiography and seismocardiography literature to date. Even standard electrophysiology recordings such as ECG and ICG exhibit motion artifacts. The methods in this paper are no different; FilterBCG and RidgeBCG cannot be used while the subject is moving, for example, during exercise. However, measurements can still be made during a rest period in the middle of an exercise, and small vibrations, such as from heavy breathing or fatigued muscles, are largely eliminated by the ensemble averaging.

## IV. CONCLUSION

In this paper two new methods to measure AVO and AVC from distal body vibrations were demonstrated. Both methods exceeded the standard method of R-J intervals in the existing literature that uses unfiltered BCG waveforms. This confirms our previous findings that R-J intervals do not correlate well to ground-truth PEP (AVO) in wearable BCG signals. In general, we recommend FilterBCG over all other methods because its AVO and AVC measurements correlated the best to an ICG ground truth in all but one case which was not statistically significant. We also believe that this is the first work to demonstrate AVC measurement from distal vibrations thus enabling measurements of LVET from distal vibrations for the first time. Future work will focus on reducing or eliminating the training step, which we believe is the main limitation of the methods we present in this paper. In particular, the training step requires an ECG and ICG, two signals that a low-cost consumer device typically would not have. Additionally, further investigation is needed to develop better cuff-less blood pressure methods that use distal vibrations as a proximal timing reference. Since accurate measurements of AVO timing from distal vibrations are now possible, and AVO is needed to measure diastolic blood pressure with aortic PTT, the efficacy of a cuff-less blood pressure device based on distal vibrations and aortic PTT can be demonstrated.

TABLE II
SUMMARY OF STATISTICAL SIGNIFICANCE

| *t*-Test[*] | p-values | |
|---|---|---|
| | AVO | AVC |
| FilterBCG vs. Filtered R-J Interval[†] | | |
| Wrist | 0.16 | N/A |
| Knee | 0.060 | N/A |
| Upper Arm | 0.53 | N/A |
| Location (FilterBCG) | | |
| Wrist vs. Upper Arm | 0.62 | 0.56 |
| Wrist vs. Knee | ++ | ++ |
| Best Axis vs. Other Axes (FilterBCG) | | |
| Wrist | ++ | ++ |
| Knee | ++ | ++ |
| Upper Arm | ++ | ++ |
| With vs. Without Filter (Upper Arm) | | |
| FilterBCG | ++ | ++ |
| Filtered R-J Interval[†] | ++ | N/A |

[*] Paired *t*-tests of the root mean square error of each subject.
[†] R-J interval does not measure AVC.
++ Indicates $p < 0.01$.

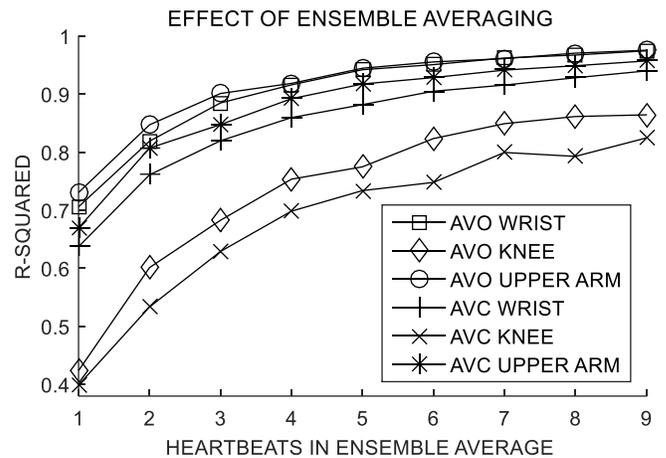

Fig. 8. Correlations of FilterBCG to ICG for several ensemble average lengths where a length of one is equivalent to no ensemble averaging. Cross validation was not used here. This plot can be used to determine a suitable ensemble average length for a given application. Measurements from the knee deteriorate more rapidly than the other locations for shorter ensemble averages.